\documentclass[iicol]{sn-jnl}
\usepackage{amsmath,amsfonts}
\usepackage{esint}
\usepackage{graphicx}
\usepackage[export]{adjustbox}
\usepackage{float}
\usepackage{cite}
\usepackage{mathtools}
\usepackage{parskip}
\usepackage{accents}
\usepackage{textcomp}
\usepackage{gensymb}
\usepackage{algorithm}
\usepackage{algpseudocode}
\usepackage[caption=false,font=normalsize,labelfont=sf,textfont=sf]{subfig}
\usepackage{textcomp}
\usepackage{stfloats}
\usepackage{url}
\usepackage{booktabs}
\usepackage{verbatim}
\def\BibTeX{{\rm B\kern-.05em{\sc i\kern-.025em b}\kern-.08em
    T\kern-.1667em\lower.7ex\hbox{E}\kern-.125emX}}
\usepackage{balance}
\usepackage{array, multirow}

\theoremstyle{thmstyleone}%
\theoremstyle{thmstyletwo}%

\theoremstyle{thmstylethree}%

\begin{document}

\title[Article Title]{Quasi-static in vivo elastography from internal displacement information only}


\author*[1]{\fnm{David G.J.} \sur{Heesterbeek}}\email{D.G.J.Heesterbeek-4@umcutrecht.nl}

\author[1]{\fnm{Max H.C.} \sur{van Riel}}\email{M.H.C.vanRiel-3@umcutrecht.nl}

\author[1]{\fnm{Ray S.S.} \sur{Sheombarsing}}\email{R.S.S.Sheombarsing@umcutrecht.nl}

\author[2]{\fnm{Tristan} \sur{van Leeuwen}}\email{T.vanLeeuwen@uu.nl}

\author[3]{\fnm{Martijn} \sur{Froeling}}\email{M.Froeling@umcutrecht.nl}

\author[1]{\fnm{Cornelis A.T.} \sur{van den Berg}}\email{C.A.T.vandenBerg@umcutrecht.nl}

\author[1]{\fnm{Alessandro} \sur{Sbrizzi}}\email{A.Sbrizzi@umcutrecht.nl}

\affil*[1]{\orgdiv{Department of Radiotherapy, Computational Imaging Group for MR Therapy and Diagnostics}, \orgname{University Medical Center Utrecht}, \orgaddress{\city{Utrecht}, \postcode{3584 CX}, \country{The Netherlands}}}

\affil[2]{\orgdiv{Department of Mathematics}, \orgname{Utrecht University}, \orgaddress{\city{Utrecht}, \postcode{3584 CD}, \country{The Netherlands}}}

\affil[3]{\orgdiv{Department of Radiology, High Field Group}, \orgname{University Medical Center Utrecht}, \orgaddress{\city{Utrecht}, \postcode{3584 CX}, \country{The Netherlands}}}

\abstract{As disease often alters the structural properties of soft tissue, noninvasive elastography techniques have emerged to quantitatively assess in vivo mechanical properties. Magnetic Resonance Elastography (MRE) based on dynamic deformations is the standard technique for imaging mechanical properties, but the viscoelastic nature of soft tissue makes the results dependent on the actuation frequency, which can be limiting. In this proof-of-principle study we propose a noise robust framework for reconstructing relative stiffness properties from quasi-static in vivo displacement fields captured on a physiological time scale. The acquisition is performed using a pneumatic pressure cuff to induce tissue deformation in a controlled manner. The reconstruction does not require boundary information which is generally hard to access in vivo nor spatial derivatives of displacement fields that are known to amplify noise. The validity of our framework is corroborated with in silico experiments on a numerical phantom. 
In vivo experiments on the thigh of a volunteer demonstrate the repeatability of the method. As an application, the quantitative change in muscle stiffness during isometric knee flexion is investigated which yielded physiologically meaningful results.}



\maketitle

\newcommand*{\skipnumber}[2][1]{%
  {\renewcommand*{\alglinenumber}[1]{}\State #2}%
  \addtocounter{ALG@line}{-#1}}

\section{Introduction}
Biomechanical properties are considered valuable clinical markers as many pathologies are related to significant structural changes. Access to these properties can provide information with regard to physiology and disease progression. Mechanical parameters are currently utilized in clinical diagnostics for evaluating chronic liver diseases\cite{glaser_review_2012}, yet their potential extends to a wide range of other applications. For example, quantitative tissue mechanical properties could allow clinicians to monitor neurological disorders like ALS and MS that are characterized by a progressive decline in muscle function \cite{kent-braun_central_2000, kent-braun_strength_1997}. The study of biomechanical properties is also relevant beyond direct clinical applications in fields such as biofabrication and regenerative medicine, where quantitative knowledge of tissue stiffness is essential\cite{guimaraes_stiffness_2020}. 

Current clinical practice for imaging mechanical tissue parameters is using Magnetic Resonance Elastography (MRE)\cite{muthupillai_magnetic_1995}. This dynamic technique requires acoustic waves to be induced in the body using an external actuator that is synchronized with the MRI machine. Information on the propagation of harmonic waves is obtained using phase contrast techniques. The acquired wave patterns can be related to the tissue's elastic properties through the conservation of momentum, represented by a hyperbolic PDE with inertia term. However, the viscoelastic nature of soft tissue causes the elastic properties to be dependent on the actuation frequency (30-200 Hz)\cite{bilston_soft_2018}, calling for methods that study these parameters on a physiological time scale. 

Unlike dynamic elastography techniques such as MRE, \mbox{(quasi-)} static elastography reconstructions use displacement fields captured on physiological time scales to encode information on the tissue's elastic properties. These reconstructions also rely on conservation of momentum. However, in the case of static elastography, the inertia term can be neglected resulting in an elliptic PDE. Pioneering work from the late '90s proposed the use of static elastography for medical imaging \cite{skovoroda_tissue_1995, chenevert_elasticity_1998, skovoroda_reconstructive_1999, steele_three-dimensional_2000}, but did not provide the robustness required for practical implementation. 


In recent years, multiple solution strategies for elasticity reconstructions from (quasi-) static displacement information have been under development, including the Finite Element Model Updating (FEMU) technique and the Virtual Fields Method (VFM). The FEMU technique\cite{chen_finite_2024, kallel_tissue_1996} is based on iteratively matching experimental data with numerical forward simulations. It is flexible and applicable in linear as well as nonlinear elastics\cite{flaschel_unsupervised_2021}, but realistic problems generally suffer from poor convergence resulting in intractable computation times. Furthermore, this approach generally requires spatially accurate boundary information for the forward simulations, which is typically not available in biomedical applications. The VFM\cite{grediac_virtual_2006} was developed in the context of experimental solid mechanics to extract material parameters from displacement information on a homogeneous sample. By applying external loads to create a heterogeneous strain field, multiple constitutive parameters can be estimated from one experiment. The method has been extended to identification of spatially dependent stiffness properties from MRI data by Avril et al. \cite{avril_3d_2008} and was tested on a simple phantom. VFM generally requires spatially accurate boundary pressure information and uses spatial derivatives of reconstructed displacement fields. These derivatives are known to amplify the noise on the measured displacement fields, causing numerical instabilities when solving the inverse problem. These requirements increase the complexity of the data acquisition process and sensitivity to noise respectively, hampering the practical application to in vivo scenarios.  

In this proof-of-principle study, we developed a novel reconstruction framework for quantitative assessment of spatially dependent stiffness properties of tissue from quasi-static displacement fields. The proposed reconstruction approach may be regarded as an extension of the VFM. Experimental data is acquired on a physiological dynamic range without the need for external actuation that is synchronized with the imaging system, as required for techniques like MRE. The main focus of our work is the noise robustness of the inverse problem for which we introduce two crucial ingredients. First, we cast the conservation of momentum equation into a weak formulation. Second, by carefully selecting the corresponding test function space, we achieve two main objectives, namely 1) we remove the need for boundary information in the reconstruction and 2) we avoid derivatives of reconstructed displacement fields. Boundary information is generally hard to access during in vivo scanning, especially for internal organs. Without additional boundary information, the reconstruction of elastic properties reduces to the task of finding the solution for the elasticity tensor in the null space of a problem matrix that we will derive. The removal of spatial derivatives from the reconstructed displacement fields, circumvents the associated noise amplification of the derivative operation, potentially increasing the noise robustness of the solution to the inverse problem.   

The validity and noise robustness of our reconstruction approach are corroborated using numerical simulations on a cylindrical coaxial domain mimicking muscle surrounding a bone. In vivo experiments are performed on the thigh of a healthy volunteer, where a controlled deformation procedure is carried out using a pneumatic pressure cuff. The displacement fields are acquired using the recently introduced Spectro-dynamic MRI framework\cite{van_riel_time-resolved_2023}, although alternative acquisition strategies, using e.g. ultrasound imaging systems can also be employed. The thigh is segmented into 3 different muscle groups based on their functionality. The quantitative stiffness values are reconstructed for each of these segments and we demonstrate the in vivo repeatability of this approach. Furthermore, we performed an experiment where the quantitative change in muscle stiffness during isometric knee flexion is investigated. The results are in agreement with what we would expect from physiology for this type of muscle activation.

\section{Model derivation}\label{sec: Theory}
We start this section by deriving the standard weak form from the Virtual Fields Method which forms the basis for our reconstruction problem (\ref{subsec: The inverse problem}). We subsequently modify it to 1) eliminate the need for boundary information and 2) to enhance the noise robustness of the solution by circumventing spatial derivatives of the measured displacement data (\ref{subsec: The proposed robust weak form}). Finally, the proposed reconstruction is obtained by discretization of the inverse problem derived in section \ref{subsec: The proposed robust weak form} using a Galerkin approximation (\ref{subsec: Discretization using the Galerkin approximation}). All vectors are denoted with $\vec{\cdot}$ unless stated otherwise, while all matrices and higher order tensors are written in \textbf{bold}. 

\subsection{The Virtual Fields Method}\label{subsec: The inverse problem}
Consider an open bounded domain $\Omega\subset\mathbb{R}^d$ with $d=2$ or $3$ and a smooth boundary embedded in a linear elastic medium. Our experiments are conducted in a quasi-static setting for which the conservation of momentum and associated boundary conditions can be written as:
\begin{align}\label{eq: conservation of momentum}
\begin{cases}
    \nabla\cdot\boldsymbol{\sigma} = \vec{0},  &\vec{x}\in\Omega,\\
    \boldsymbol{\sigma}\cdot\hat{n}=\vec{p}, &\vec{x}\in\partial\Omega, 
\end{cases}     
\end{align}
where $\boldsymbol{\sigma}$ is the Cauchy stress tensor, $\vec{p}$ the pressure, $\partial\Omega$ the boundary of domain $\Omega$ and $\hat{n}$ the outward unit normal vector to the boundary. To arrive at the weak form, we rewrite the strong form in \eqref{eq: conservation of momentum} to:
\begin{equation}\label{eq: weak_form}
    \int_{\Omega}\vec{\eta}\cdot(\nabla\cdot\boldsymbol{\sigma})\text{d}V = 0,
\end{equation}
where $\vec{\eta}$ is an arbitrary vector-valued test function. In this paper, we assume a linear relation between the deformations (strain) and the resulting forces (stress) such that the following constitutive equation holds: $\boldsymbol{\sigma} = \boldsymbol{C}:\boldsymbol{\epsilon}$. Here $\boldsymbol{C}$ is the fourth order elasticity tensor that describes the mechanical parameters and $\boldsymbol{\epsilon}$ the second order linear strain tensor. The linear strain tensor is defined as $\boldsymbol{\epsilon} := \frac{1}{2}(\nabla\vec{u} + (\nabla\vec{u})^T)$ with $\vec{u}$ the displacement field. Using integration by parts, the boundary condition from \eqref{eq: conservation of momentum} and the fundamental symmetries in $\boldsymbol{C}$, we can rewrite this expression into the general weak formulation for $\vec{u}$: 
\begin{equation}\label{eq: VFM_u}
    \int_{\Omega} \nabla\vec{\eta}:(\boldsymbol{C}:\nabla\vec{u})\text{d}V = \int_{\partial \Omega} \vec{\eta}\cdot\vec{p}\text{d}A,
\end{equation}
as derived in the Supplementary materials 7.1. Here $``:"$ is the double dot operator that contracts the last two indices of the first tensor with the first two indices of the second tensor. This equation couples the elastic properties with the material displacements. It is the standard form used in the VFM, where $\vec{\eta}$ acts as the `virtual displacement'. 

\subsection{The proposed noise robust inversion problem}\label{subsec: The proposed robust weak form}
The inverse problem of recovering $\boldsymbol{C}$ from measurements of $\vec{u}$ and $\vec{p}$ using \eqref{eq: VFM_u} is impractical for reasons described in the introduction: it requires inaccessible pressure information on the boundaries and spatial derivatives of the measured displacement fields, that amplify the noise.

To remove the need for boundary pressure, we restrict the space of test functions $\vec{\eta}$ to only include functions with the property $\vec{\eta}|_{\partial\Omega} = \vec{0}$. This restriction causes the right hand side of \eqref{eq: VFM_u} to disappear resulting in the following equation: 
\begin{equation}\label{eq: FEM}
    \int_{\Omega} \nabla\vec{\eta}:(\boldsymbol{C}:\nabla\vec{u})\text{d}V = 0.
\end{equation}
To remove the need for spatial derivatives on reconstructed displacement data $\vec{u}$, we use integration by parts a second time, transferring the derivative on the measured field $\vec{u}$ to the analytically known field $\vec{\eta}$. This approach was inspired by Reinbold et al.\cite{reinbold_using_2020}, who used this transfer of derivatives for data driven learning of PDEs with constant coefficients. For the purpose of brevity, we use the Einstein summation convention for the remainder of this discussion. If we use integration by parts, \eqref{eq: FEM} becomes: 
\begin{equation}
\begin{aligned}\label{eq: VFM_A}
    &\int_{\Omega} (\partial_j\eta_iC_{ijkl})\partial_lu_k\text{d}V = \\
    &\int_{\partial\Omega} (\partial_j\eta_iC_{ijkl})u_k\hat{n}_l\text{d}A - \int_{\Omega}\partial_l(\partial_j\eta_iC_{ijkl})u_k\text{d}V = 0.
\end{aligned}
\end{equation}
Evaluating the newly introduced boundary integral is challenging in practice due to the complexity of representing spatial discontinuities in the displacement field at the tissue-air interface. To circumvent the use of the displacement fields at the boundary, we further restrict the test function space to include only functions with the additional property $\nabla\vec{\eta}|_{\partial\Omega} = \boldsymbol{0}$. This results in the boundary integral reducing to zero such that we arrive at the following necessary condition on $\boldsymbol{C}$: 
\begin{equation}\label{eq: robust_weak}
    \int_{\Omega}\partial_l(\partial_j\eta_i C_{ijkl})u_k\text{d}V = 0.
\end{equation}
The inverse problem arises from this equation for the restricted test function space. In \eqref{eq: robust_weak}, we see $\vec{u}$ as the known parameter and $\boldsymbol{C}$ as the unknown variable of interest. More formally, we define the following bilinear map $B_{\vec{u}}:W^{1,\infty}(\Omega, \boldsymbol{T}^4_{\text{sym}})\times\mathcal{V}\rightarrow\mathbb{R}$ by: 
\begin{equation}
    B_{\vec{u}}(\boldsymbol{C}, \vec{\eta}) := \int_{\Omega}\partial_l(\partial_j\eta_i C_{ijkl})u_k\text{d}V, 
\end{equation}
with $W^{k,p}$ the standard Sobolev space of $k$-times weakly differentiable functions with finite $L^p$ norm, $\boldsymbol{T}^4_{\text{sym}}$ the space of all tensors with the symmetry properties described in (S.8) of the Supplementary materials and $\mathcal{V}:= \{\vec{\eta}\in W^{2,2}(\Omega, \mathbb{R}^d):\vec{\eta}|_{\partial\Omega}=\vec{0}, \nabla\vec{\eta}|_{\partial\Omega}=\boldsymbol{0}\}$.
Finally, we define our inverse problem as follows: given $\vec{u}\in L^2(\Omega, \mathbb{R}^d)$, 
\begin{flalign}\label{eq: inverse_problem}
    \text{find } \boldsymbol{C} & \text{ s.t.} &  B_{\vec{u}}(\boldsymbol{C}, \vec{\eta})= 0,  &&& \forall\vec{\eta}\in \mathcal{V}.
\end{flalign}
Note that in this formulation of the inverse problem, we do not require boundary pressure information $\vec{p}$ nor derivatives of the measured displacement field $\vec{u}$.

\subsection{Discretization using the Galerkin approximation}\label{subsec: Discretization using the Galerkin approximation}
\begin{figure*}[t]
    \centering
    \includegraphics[width=\linewidth]{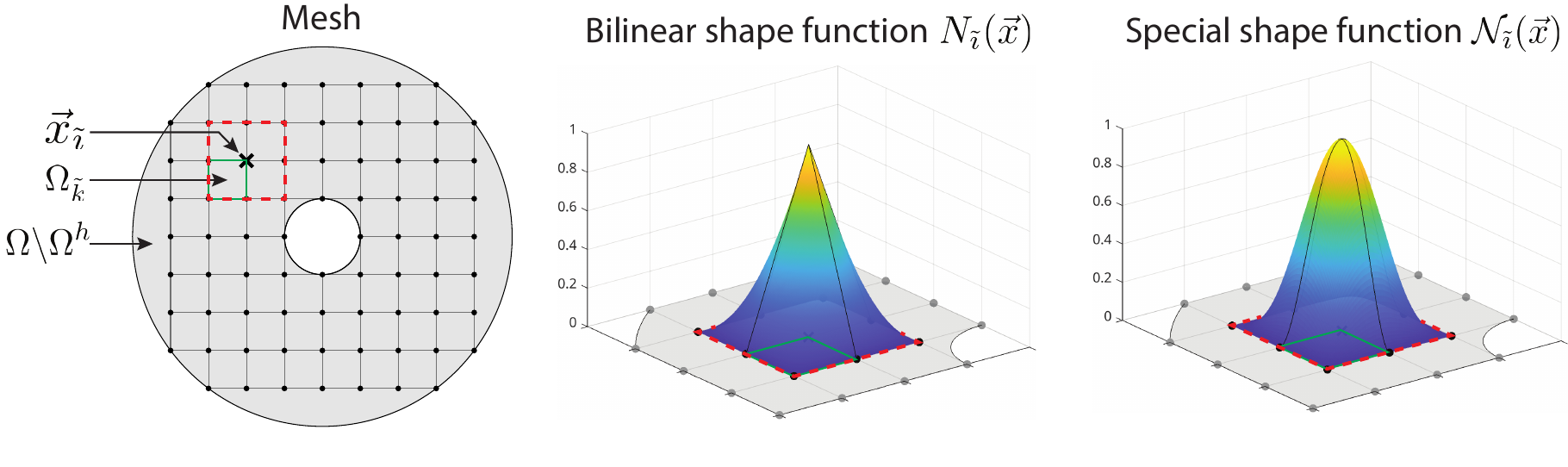}
    \caption{(Left) Schematic depiction of the coaxial domain mimicking muscle surrounding a bone discretized using quadrilaterals. The mesh is chosen extremely coarse for illustrative purposes. (Middle) The bilinear shape function $N_{\tilde{\imath}}(\vec{x})$ in the neighborhood of $\vec{x}_{\tilde{\imath}}$. The family of bilinear shape functions $N(\vec{x})$ around all nodes form a basis for the solution $\boldsymbol{C}(\vec{x})$. (Right) The special shape function $\mathcal{N}_{\tilde{\imath}}(\vec{x})$ in the neighborhood of $\vec{x}_{\tilde{\imath}}$. The family of special shape functions $\mathcal{N}(\vec{x})$ around all internal nodes form a basis for the test functions $\vec{\eta}(\vec{x})$.}
    \label{fig: grid}
\end{figure*}
Now that we have set up a more noise robust inverse problem, we can start constructing approximate solutions by discretizing the operator $B_{\vec{u}}(\boldsymbol{C}, \vec{\eta})$. To this end, the domain $\Omega$ is subdivided into $n_{el}$ elements $\Omega_{\tilde{k}}$, using $n_{node}$ nodes placed at $\vec{x}_{\tilde{\imath}}$. The tildes on the indices are used to distinguish them from the indices used to derive the noise robust inverse problem in Section \ref{subsec: The proposed robust weak form}. The discretized domain is denoted by $\Omega^h$ and we assume that $\Omega^h=\bigcup\limits_{\tilde{k}=1}^{n_{el}}\Omega_{\tilde{k}}\approx\Omega$. This work focuses on the case $d=2$, but extending the developed framework to 3 dimensions is straightforward. For the subdivision, quadrilaterals are adopted for simplicity, but this can be generalized to different element shapes as well. 

The spatially dependent elasticity tensor $\boldsymbol{C}(\vec{x})$ is expanded using a standard basis of local bilinear shape functions (see Fig. \ref{fig: grid}). Each node $\tilde{\imath}$ has a corresponding shape function $N_{\tilde{\imath}}(\vec{x})$ that has the property of being equal to one in the corresponding node and vanishes in all other nodes: $N_{\tilde{\imath}}(\vec{x}_{\tilde{\imath}})=\delta_{\tilde{\imath}\tilde{\jmath}}$ with $\delta_{\tilde{\imath}\tilde{\jmath}}$ the Kronecker delta. The elasticity tensor $\boldsymbol{C}(\vec{x})$ can now be approximated using the bilinear shape functions located around the $n_{node}$ nodes. Storing the bilinear shape functions $N_{\tilde{\imath}}(\vec{x})$ and the corresponding nodal elasticity tensors $\boldsymbol{C}_{\tilde{\imath}}$ using a compact column notation (denoted with a tilde below the symbol) as follows: 
\begin{equation}
\begin{aligned}
    \underaccent{\tilde}{N}(\vec{x}) &= \left[N_1(\vec{x}), N_2(\vec{x}), N_3(\vec{x}), ..., N_{n_{node}}(\vec{x})\right]^T,\\
    \underaccent{\tilde}{\boldsymbol{C}} &= \left[\boldsymbol{C}_1, \boldsymbol{C}_2, \boldsymbol{C}_3, ..., \boldsymbol{C}_{n_{node}}\right]^T,
\end{aligned}
\end{equation}
we can write the Galerkin approximation as follows: 
\begin{equation}
    \boldsymbol{C}(\vec{x}) = \underaccent{\tilde}{N}^T(\vec{x})\underaccent{\tilde}{\boldsymbol{C}}.  
\end{equation}

The spatially dependent arbitrary test function $\vec{\eta}(\vec{x})$ is expanded using a local basis of "special shape functions" (see Fig. \ref{fig: grid}). For an ensemble of 4 connected quadrilateral elements $\{(x,y): |x - x_{\tilde{\imath}}|\leq H_x, |y - y_{\tilde{\imath}}|\leq H_y\}$ that is centered around node $\tilde{\imath}$ located at $\vec{x}_{\tilde{\imath}}=(x_{\tilde{\imath}},y_{\tilde{\imath}})$ in the interior (red dashed square in Fig. \ref{fig: grid}), the special shape function is defined as:
\begin{equation}
    \mathcal{N}_{\tilde{\imath}}(\vec{x}) = \left(\left(\frac{x-x_{\tilde{\imath}}}{H_x}\right)^2-1\right)^2\left(\left(\frac{y-y_{\tilde{\imath}}}{H_y}\right)^2-1\right)^2.
\end{equation}
This family of polynomial functions satisfies the requirements for the test functions outlined in section \ref{subsec: The proposed robust weak form}, such that an expansion in this local basis ensures $\vec{\eta}\in\mathcal{V}$. The test function $\vec{\eta}(\vec{x})$ can now be described using the special shape functions located around the $n_{node}^{int}$ interior nodes. The nodes at the boundary are left out as the test function is supposed to be 0 here. This results in the following local functions and associated weights: 
\begin{equation}
\begin{aligned}
    \underaccent{\tilde}{\mathcal{N}}(\vec{x}) &= \left[\mathcal{N}_1(\vec{x}), \mathcal{N}_2(\vec{x}), \mathcal{N}_3(\vec{x}), ..., \mathcal{N}_{n_{node}^{int}}(\vec{x})\right]^T,\\
    \underaccent{\tilde}{\vec{\eta}} &= \left[\vec{\eta}_1, \vec{\eta}_2, \vec{\eta}_3, ..., \vec{\eta}_{n_{node}^{int}}\right]^T,
\end{aligned}
\end{equation}
such that the test functions can be approximated using a finite dimensional expansion as: 
\begin{equation}
    \vec{\eta}(\vec{x}) = \underaccent{\tilde}{\mathcal{N}}^T(\vec{x})\underaccent{\tilde}{\vec{\eta}}.
\end{equation}
The subspace spanned by the special shape functions satisfies the requirements for the test function such that $\vec{\eta}\in \mathcal{V}$. For the Jacobian of $\vec{\eta}$ we use:  
\begin{equation}
    \partial_j\eta_i(\vec{x}) = \left(\partial_j\underaccent{\tilde}{\mathcal{N}}(\vec{x})\right)^T\underaccent{\tilde}{\eta}_i,
\end{equation}
where
\begin{equation}
  \partial_j\underaccent{\tilde}{\mathcal{N}}(\vec{x}) = \left[\partial_j\mathcal{N}_1(\vec{x}), \partial_j\mathcal{N}_2(\vec{x}), \partial_j\mathcal{N}_3(\vec{x}), ..., \partial_j\mathcal{N}_{n_{node}^{int}}(\vec{x})\right]^T, 
\end{equation}
and analogously for the Hessian of $\vec{\eta}$ and the Jacobian of $\boldsymbol{C}$. 
Now that all relevant spatially dependent functions are expanded into the basis functions described above and the domain is subdivided into elements, \eqref{eq: robust_weak} can be represented as follows:
\begin{equation}
    \underaccent{\tilde}{\eta}^T_i\int_{\Omega^h}\partial_l(\partial_j\underaccent{\tilde}{\mathcal{N}} \underaccent{\tilde}{N}^T)u_k\text{d}V\underaccent{\tilde}{C}_{ijkl} = 0.  
\end{equation}
Using the product rule for differentiation on the $\partial_l$ operator yields: 
\begin{equation}\label{eq: Robust_FEM1}
    \underaccent{\tilde}{\eta}^T_i\int_{\Omega^h}\left(\partial_l\partial_j\underaccent{\tilde}{\mathcal{N}} \underaccent{\tilde}{N}^T + \partial_j\underaccent{\tilde}{\mathcal{N}}\partial_l\underaccent{\tilde}{N}^T\right)u_k\text{d}V\underaccent{\tilde}{C}_{ijkl} = 0.  
\end{equation}
Note that $u_k=u_k(\vec{x})$ is assumed to be known continuously from measurements and interpolation. 

At this stage we should recall that \eqref{eq: Robust_FEM1} holds for \emph{any} test function $\vec{\eta}$ that can be built up in the basis of special shape functions described above. This is only possible if and only if: 
\begin{equation}\label{eq: robust_FEM2}
    \int_{\Omega^h}\left(\partial_l\partial_j\underaccent{\tilde}{\mathcal{N}} \underaccent{\tilde}{N}^T + \partial_j\underaccent{\tilde}{\mathcal{N}}\partial_l\underaccent{\tilde}{N}^T\right)u_k\text{d}V\underaccent{\tilde}{C}_{ijkl} = 0,
\end{equation}
for all $i=\{1...d\}$. We have effectively rewritten our inverse problem as a matrix-vector product equation, where the non-trivial solution for the coefficients $\underaccent{\tilde}{C}_{ijkl}$ determines the values of the elasticity tensor at the imposed grid. Finally, we note that solution to our inverse problem will be a non-trivial solution of the homogeneous system determined by the reconstructed displacements: 
\begin{equation}\label{eq: Inverse problem}
    \boldsymbol{A}\left(\vec{u}\right)\vec{c} = \vec{0}, 
\end{equation}
where $\boldsymbol{A}$ is the problem matrix containing the evaluation of the integral $\int_{\Omega^h}\left(\partial_l\partial_j\underaccent{\tilde}{\mathcal{N}} \underaccent{\tilde}{N}^T + \partial_j\underaccent{\tilde}{\mathcal{N}}\partial_l\underaccent{\tilde}{N}^T\right)u_k \text{d}V$ and $\vec{c}$ the vector containing the coefficients $\underaccent{\tilde}{C}_{ijkl}$ that describe the elastic parameters we are looking for. To evaluate the matrix entries, numerical evaluation using Gaussian quadrature is used. In practice, the omission of boundary pressure information means that we can only reconstruct the material properties up to an arbitrary global scaling, yielding relative stiffness values rather than absolute ones.

\section{Methods}\label{sec: Methods}
To reconstruct the elastic parameters, a solution strategy was developed using the necessary condition on $\vec{c}$ defined in \eqref{eq: Inverse problem}. Additionally, we explain how assumptions on region-wise homogeneity of mechanical properties can be used to enhance the robustness (\ref{subsec: reconstruction}). In silico as well as in vivo experiments were conducted to test the proposed reconstruction framework (\ref{subsec: Experimental set-up}). All reconstructions were performed for two-dimensional domains ($d=2$) using the plane strain approximation (zero strain in the third dimension). 

\subsection{Reconstruction of elasticity parameters}\label{subsec: reconstruction}
Eliminating the boundary pressure information from \eqref{eq: VFM_u} by restricting the test function space, necessarily renders the problem matrix $\boldsymbol{A}$ rank-deficient. To arrive at an inverse problem that is uniquely solvable up to a global multiplicative constant, four linearly independent displacement fields are required\cite{barbone_elastic_2004}. When $n_{exp}$ displacement fields $\vec{u}^{(m)}$ are available from applying different boundary conditions to a single object, these fields can be leveraged to improve the conditioning of the reconstruction problem as follows: 
\begin{equation}\label{eq: Inverse problem multi}
\boldsymbol{\mathcal{A}}\vec{c} := 
   \begin{bmatrix}
       \boldsymbol{A}\left(\vec{u}^{(1)}\right)\\
       \boldsymbol{A}\left(\vec{u}^{(2)}\right)\\
       \vdots \\
       \boldsymbol{A}\left(\vec{u}^{(n_{exp})}\right)
   \end{bmatrix}
   \vec{c} = \vec{0},
\end{equation}
with $\boldsymbol{\mathcal{A}}$ the total problem matrix. 
One way of finding a non-trivial solution to \eqref{eq: Inverse problem multi} is by posing the problem as a constrained $L^2$-minimization: 
\begin{align}\label{eq: minprob}
\begin{split}
    \underset{\vec{c}}{\arg\min}&~\|\boldsymbol{\mathcal{A}}\vec{c}\|_2\\
    \text{subject to:}&~~\|\vec{c}\|_1 = \alpha,
\end{split}
\end{align}
with $\alpha$ an arbitrary constant, since the constraint on $\vec{c}$ is imposed solely to guarantee a non-trivial solution. 

To solve the minimization problem in \eqref{eq: minprob}, the form of the elasticity tensor $\boldsymbol{C}$ should be specified. We consider a linear isotropic model characterized by two elastic parameters: the Poisson ratio ($\nu$), which measures the material's compressibility, and the Young's modulus ($E$), which quantifies the material's stiffness. The Poisson ratio is assumed to be homogeneous and known. For the in silico experiments a Poisson ratio of $\nu=0.4$ was used. For the in vivo experiments we consider soft tissue to be an incompressible material, i.e. the Poisson ratio is equal to $0.5$. However, for the numerical stability of our reconstructions we used $\nu=0.49$. The elasticity tensor for a linear isotropic material has the following form: 
\begin{equation}
    C_{ijkl}(\vec{x}) = \frac{\nu E(\vec{x})}{(1+\nu)(1-2\nu)}\delta_{ij}\delta_{kl} + \frac{E(\vec{x})}{1+\nu}\delta_{ik}\delta_{jl}. 
\end{equation}
The number of unknowns in the vector $\vec{c}$ depends on the number of elastic parameters in the chosen model ($n_{param}$) and the number of nodes in the spatial discretization ($n_{node})$. With the assumptions outlined above, we have $n_{param}=1$ such that our inverse problem effectively reduces to finding the spatially dependent Young's modulus $E(\vec{x})$. The dimension of the total problem matrix $\boldsymbol{\mathcal{A}}$ is $n_{node}^{int}n_{exp}d\times n_{node}n_{param}$. The extension to other linear models with different elastic parameters (transversely isotropic, orthotropic, anisotropic) is straightforward, but will increase the number of unknowns as $n_{param}$ grows, affecting the conditioning of the problem. We leave these model extensions for future work.  

Solving \eqref{eq: minprob} can be challenging for local, voxel-wise mappings of $\boldsymbol{C}$. To improve the conditioning of the problem, we might look at a simplification where material properties are considered region-wise constant. In practice, we assume that individual anatomical structures (3 different muscles groups from the thigh in this work) have approximately homogeneous properties. By segmenting the spatial domain in $n_{seg}$ segments according to the anatomy, we can group unknown variables and simplify the problem to reconstruction of mechanical properties for a whole structure rather than for every location. Incorporating this approximation reduces the number of unknowns from $n_{node}n_{param}$ to $n_{seg}n_{param}$. 

The local test functions that are created around a node, cover 4 elements (see Fig. \ref{fig: grid}). We combine multiple local test functions associated with different nodes, to cover an entire segment. This approach was also used in the Virtual Fields Method for non-homogeneous media in \cite{mei_improving_2019}.

\subsection{Experimental set-up}\label{subsec: Experimental set-up}
\begin{figure}[t]
    \centering
    \includegraphics[width=\linewidth]{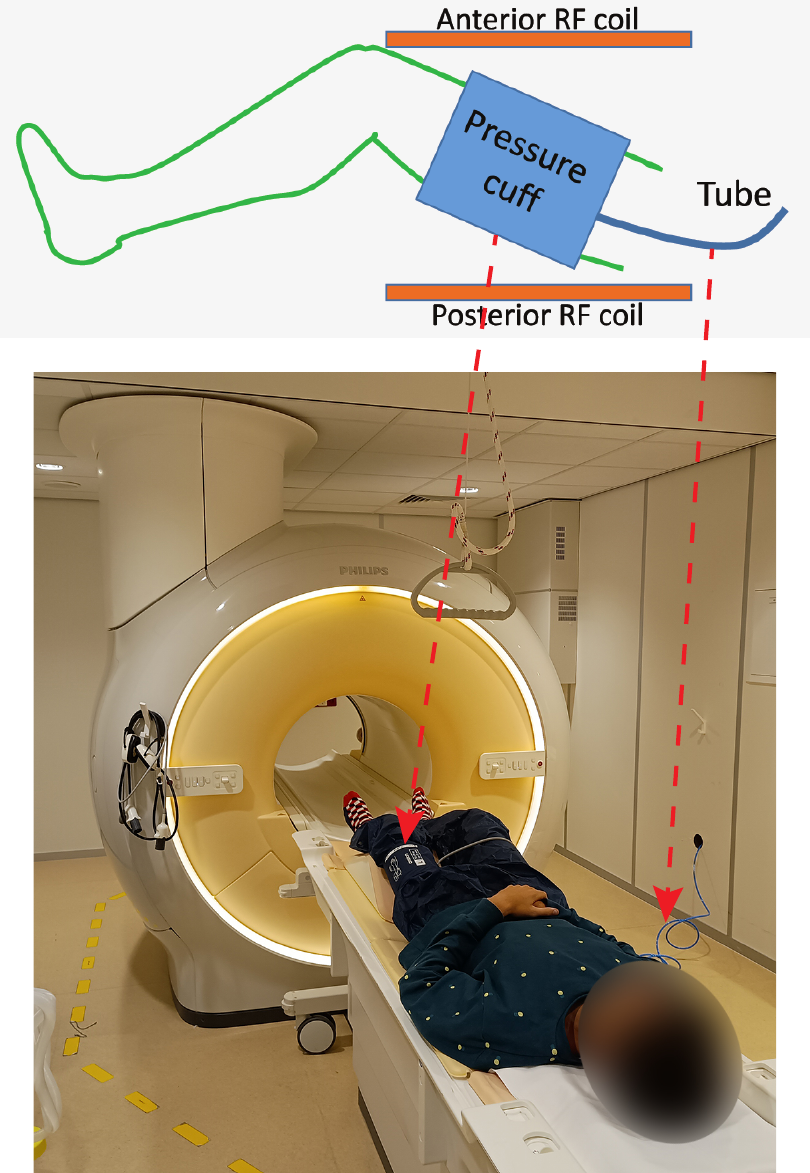}
    \caption{(Top) Schematic representation of the set-up with the pneumatic pressure cuff and coil placement. (Bottom) Volunteer in supine position on the scanner table (anterior RF coil not shown).}
    \label{fig: in-vivo}
\end{figure}
In silico tests on a numerical phantom were conducted to validate the proposed reconstruction approach through two distinct experiments: one without a region-based mechanical uniformity assumption (1.1) and one with this assumption (1.2). For the in vivo proof-of-principle results, we again carried out two experiments: one to indicate the repeatability (2.1) and one application study where the stiffness values during isometric knee flexion were investigated and compared to the muscle in relaxed state (2.2). The reconstruction procedure is implemented in Matlab (The MathWorks Inc., Natick, MA, USA).  

\subsubsection{In silico experiments}
\begin{figure*}[t!]
    \centering
    \includegraphics[width=\linewidth]{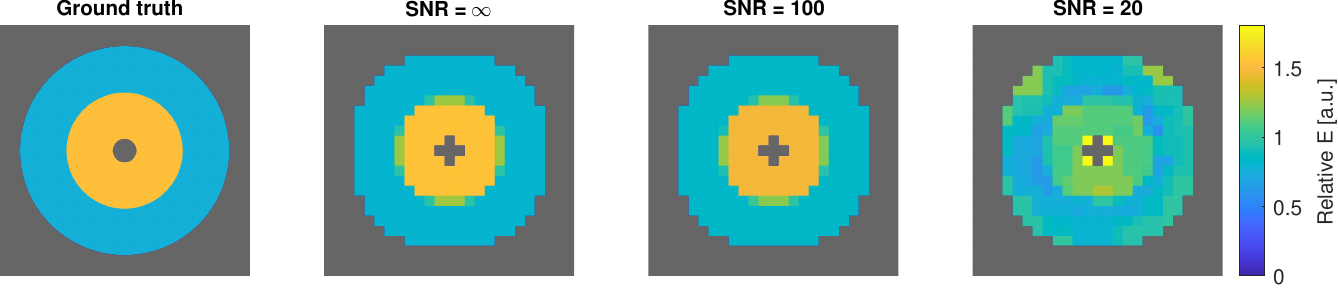}
    \caption{In silico Experiment 1.1. On the domain, two materials types are present ($n_{seg}=2$) with different Young's moduli (inner ring: $30$kPa, outer ring: $15$kPa). The relative Young's moduli are reconstructed for each spatial location such that the sum of all nodes equals $\alpha=n_{node}$.}
    \label{fig: InSilicoLoc}
\end{figure*}

\begin{figure}[b!]
    \centering
    \includegraphics[width=\linewidth]{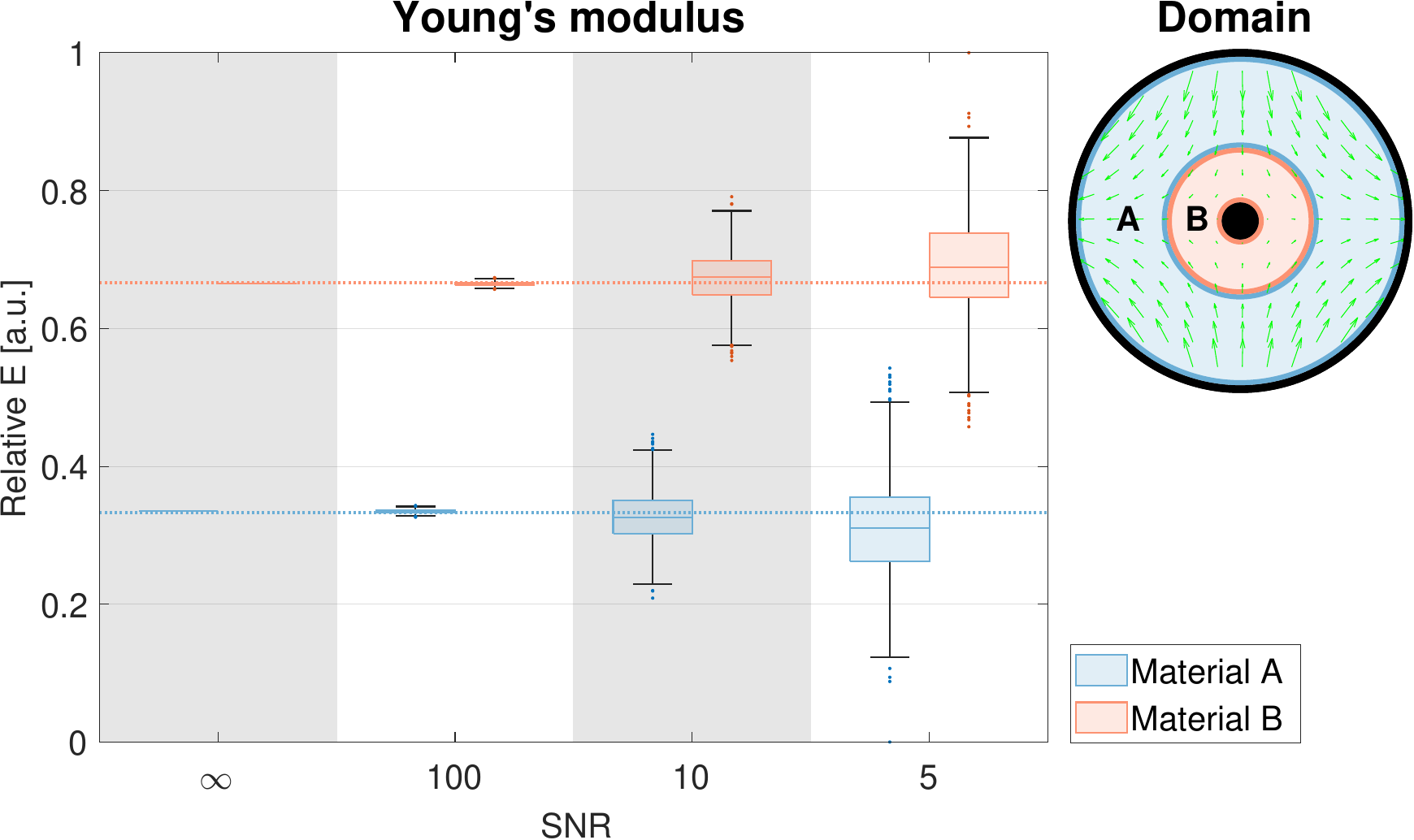}
    \caption{In silico Experiment 1.2. The dashed lines in the box plot are the ground truth relative Young's moduli. The green arrows in the domain represent one displacement field (out of $8$ used for reconstruction) and the black circle in the middle the bone structure. On the domain, two material types are present ($n_{seg}=2$) with different Young's moduli (material A: $15$kPa, material B: $30$kPa). The relative Young's moduli are reconstructed for the two segments such that their sum equals $\alpha=1$.}
    \label{fig: InSilico1}
\end{figure}

\begin{figure}[b!]
    \centering
    \includegraphics[width=\linewidth]{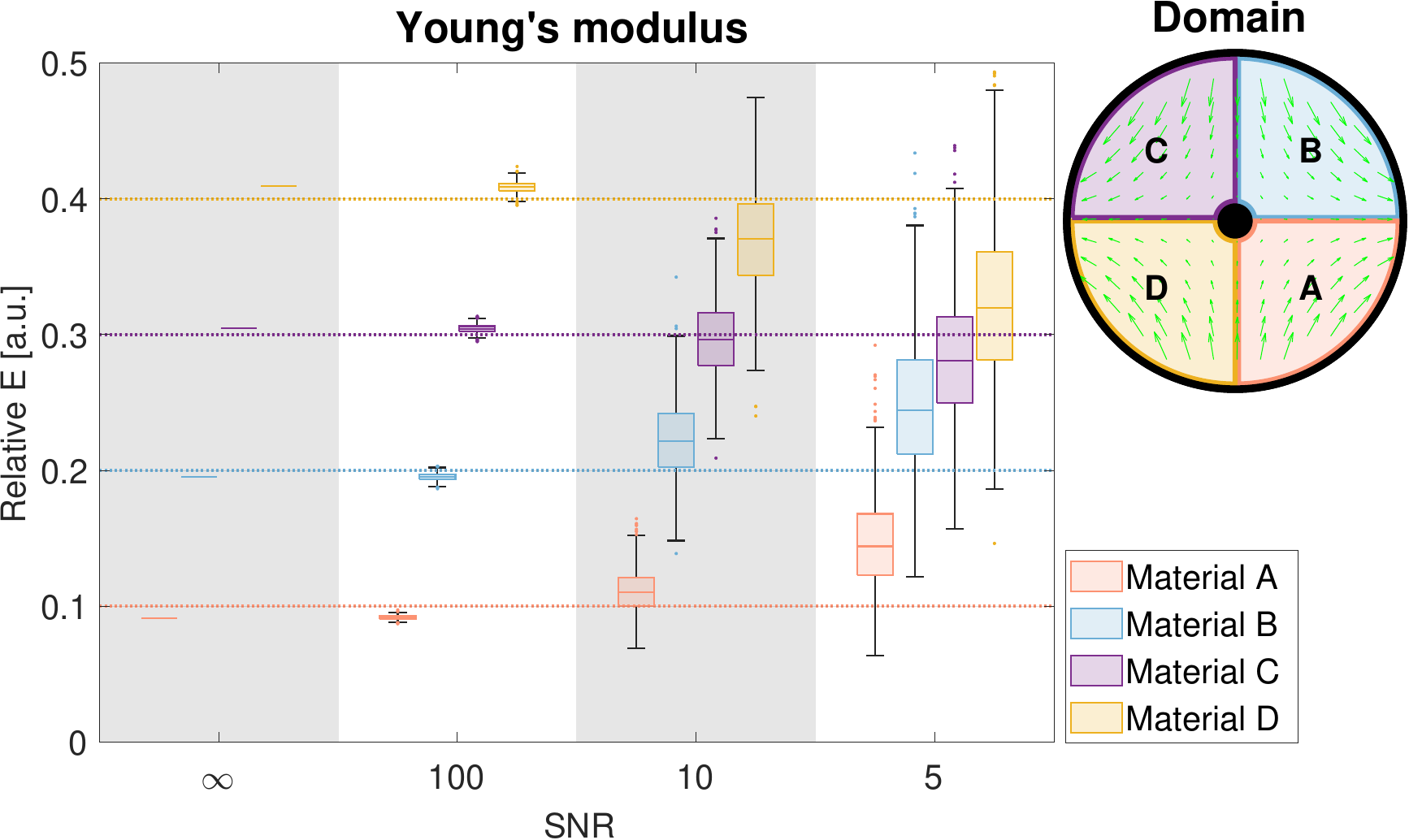}
    \caption{In silico Experiment 1.2. The dashed lines in the box plot are the ground truth relative Young's moduli. The green arrows in the domain represent one displacement field (out of $8$ used for reconstruction) and the black circle in the middle the bone structure. On the domain, four materials are present ($n_{seg}=4$) with different Young's moduli (material A: $10$kPa, material B: $20$kPa, material C: $30$kPa, material D: $40$kPa). The relative Young's moduli are reconstructed for the four segments such that their sum equals $\alpha=1$.}
    \label{fig: InSilico4}
\end{figure}

To validate the proposed reconstruction approach, $n_{exp}=8$ ground truth displacement fields were generated on a cylindrical domain mimicking muscle surrounding a bone under 2D plane strain conditions. These displacement fields were obtained by applying distinct boundary conditions to the outer rim, while fixing the inner rim of the domain for a single elastic configuration. All simulations are performed using the finite element software FEBio\cite{maas_febio_2012}. A linear isotropic model is used for the forward simulations with a Poisson ratio of $\nu=0.4$ which is assumed to be known as mentioned in section \ref{subsec: reconstruction}. The in silico validation is performed on two different heterogeneous material configurations which exhibit discontinuities ($\boldsymbol{C}\in L^{\infty}(\Omega, \boldsymbol{T}^4_{\text{sym}})$). The domains contain $n_{seg}=2$ types of material (see Figs. \ref{fig: InSilicoLoc} and \ref{fig: InSilico1}) or $n_{seg}=4$ types of material (see Fig. \ref{fig: InSilico4}) or "muscle groups" for which the elastic properties are homogeneous. The robustness of the proposed reconstruction methods is evaluated for different noise levels. The measured displacement fields on the grid are defined as $\vec{u} = \vec{u}_{\text{GT}} + \vec{e}$, where $\vec{u}_{\text{GT}}$ is the ground truth displacement field and $\vec{e}\sim N(\vec{0},\sigma\mathbb{I})$ the added Gaussian perturbation. We define the signal-to-noise ratio as: 
\begin{equation}
\text{SNR} = \frac{\text{mean}(|\vec{u}|)}{\sqrt{2}\sigma},
\end{equation}
where the factor $\sqrt{2}$ naturally arises in the 2D context.

\emph{Experiment 1.1: local, voxel-wise stiffness characterization}~\\
For Experiment 1.1, we reconstructed the spatial distribution of the relative Young's modulus in its full generality (no region-wise homogeneity assumption is used) by solving: 
\begin{align}\label{eq: minprobLoc}
\begin{split}
    \underset{\vec{c}}{\arg\min}&~\|\boldsymbol{\mathcal{A}}\vec{c}\|_2 + \lambda \text{TV}(\vec{c})\\
    \text{subject to:}&~~\|\vec{c}\|_1 = \alpha.
\end{split}
\end{align}
Here $\text{TV}$ is the total variation norm and $\lambda$ the related regularization parameter. For this experiment $\alpha=n_{node}$. The noise robustness was investigated for $\text{SNR}=\infty, 100, 20$ for which the regularization parameter was set to $\lambda=5\cdot10^{-5}, 5\cdot10^{-5}, 10^{-5}$ respectively selected using a heuristic approach. The dimensions of the input displacement fields are $193\times193$ and the dimensions of the solution for the elastic parameters are $25\times25$. This problem is solved using \texttt{CVX}, a Matlab package for specifying and solving convex problems \cite{cvx_research_inc_cvx_2012, m_grant_graph_2008}. Due to the anticipated lower noise robustness of this fully localized reconstruction compared to the region-wise approach described in the following paragraph, we opted not to apply this method for the in vivo data.

\emph{Experiment 1.2: region-wise stiffness characterization}~\\
For Experiment 1.2, a region-wise homogeneity assumption based on segmentation information is incorporated into the problem to enhance its noise robustness. For this experiment $\alpha=1$. The noise robustness is investigated for $\text{SNR}=\infty, 100, 10, 5$ and the error analysis for this experiment is performed using a Monte Carlo simulation with $1000$ noise realizations for every evaluated SNR value (except $\infty$). The dimensions of the input displacement fields are $96\times96$. The relative Young's moduli of the segments are reconstructed by solving \eqref{eq: minprob} using the \texttt{lsqnonlin} function in Matlab.

\begin{figure}[t]
    \centering
    \includegraphics[width=\linewidth]{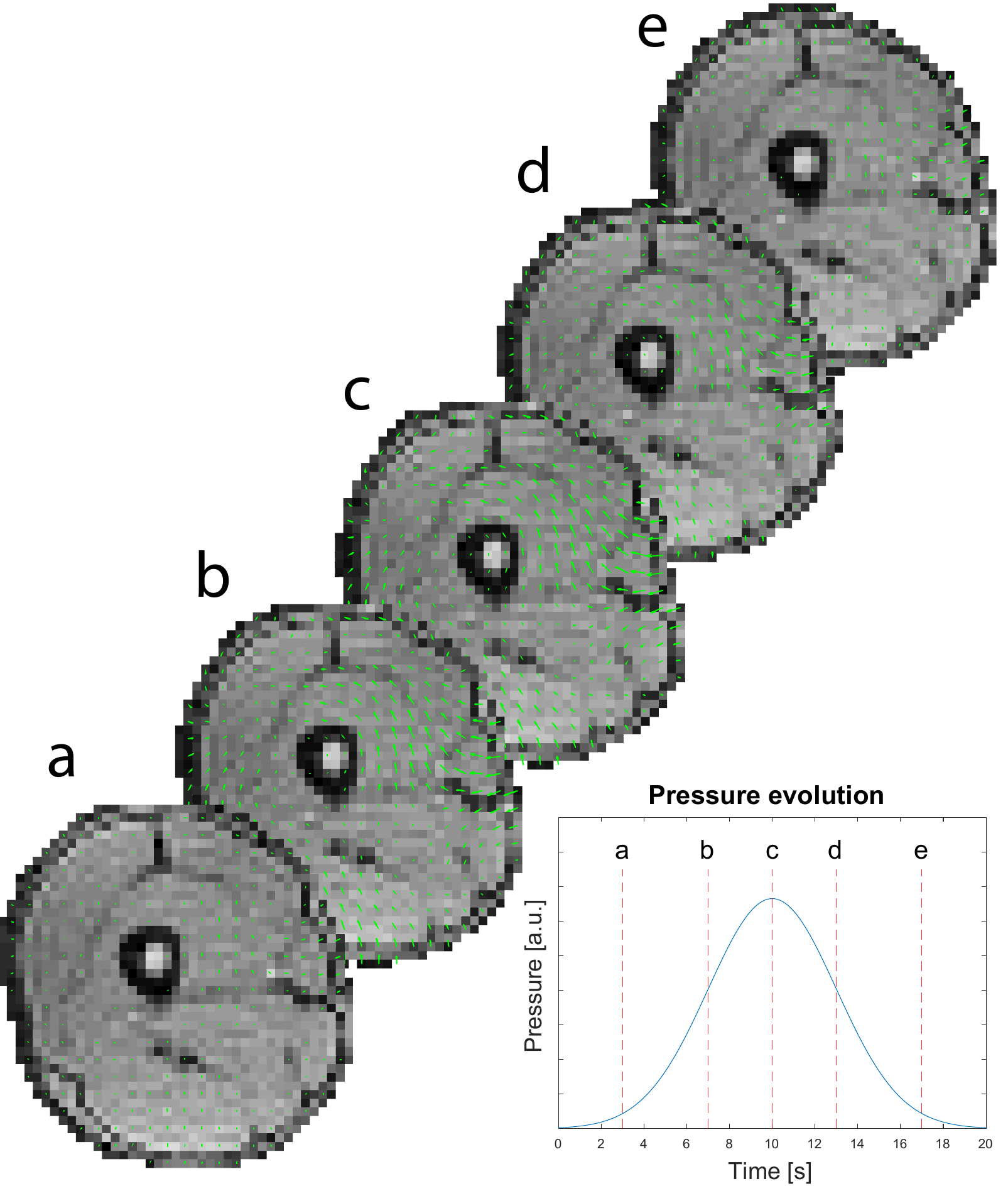}
    \caption{Displacement fields (green arrows) at different time points during inflation and deflation of the pressure cuff. The full video with a spatial resolution of $3.5$mm isotropic and a temporal resolution of $345.6$ms can be found in the Supplementary materials.}
    \label{fig: displacement}
\end{figure}

\subsubsection{In vivo experiments}
For the in vivo experiments, an MRI-compatible protocol was developed to induce repeatable deformations in the thigh muscle. This was achieved using a pneumatic pressure cuff with pressure readings given by a digital manometer (M6 Comfort AFib, Omron Healthcare Co., Kyoto, Japan) as shown in Fig. \ref{fig: in-vivo}. The protocol was applied to a healthy volunteer who provided informed consent. 
The volunteer is continuously scanned during inflation and deflation of the cuff, resulting in time-dependent displacement fields (see Fig. \ref{fig: displacement}). The reconstruction of these displacement fields $\vec{u}(x,y,t)$ is performed using the Spectro-dynamic MRI framework which is able to distill displacement fields and images with a high spatiotemporal resolution, directly from $k$-space data acquired by the MRI scanner. The method uses a motion model that exploits the spatiotemporal correlations translated to the $k$-space domain\cite{van_riel_time-resolved_2023}. The time dependent displacement fields $\vec{u}(x,y,t)$ serve as input for our inverse problem. As noted in section \ref{subsec: reconstruction}, the unique reconstruction of elastic properties requires multiple distinct displacement fields from the same object. Each individual time point $t_i$ is considered a distinct displacement field that provides new information on the underlying stiffness parameters. 

The data is acquired on a 1.5T Philips Ingenia MRI scanner (Philips, Best, The Netherlands). We used a 3D spoiled gradient echo sequence with $\text{TR}=5.4\text{ms}$, $\text{TE}=2.3\text{ms}$ and a flip angle of $6^{\circ}$. The field of view was set to be $448\text{mm}\times224\text{mm}\times224\text{mm}$, with $3.5\text{mm}$ isotropic resolution. We reconstructed the time-resolved displacement field $\vec{u}(x,y,t)$ with a temporal resolution of $345.6$ms during inflation and deflation of the pressure cuff which is considered high enough to capture the motion reliably. The $k$-space sampling trajectory was played out using a Cartesian Acquisition with Spiral PRofile (CASPR) order to efficiently capture the motion \cite{prieto_highly_2015}. The read-out is performed along the left-right direction in the patient reference frame. The posterior and anterior coils were used for data acquisition as schematically visualized in Fig. \ref{fig: in-vivo}. A set of 61 time frames of a 2D slice with minimal through plane motion was chosen as input for the proposed reconstruction framework. These thigh images were manually segmented into $3$ muscle groups based on functionality: the anterior, medial and posterior compartment (see Figs. \ref{fig: Reproducibility} and \ref{fig: Comparison}) that are approximated to have homogeneous elastic properties.

\emph{Experiment 2.1: in vivo elastography and repeatability}~\\
The in vivo repeatability is assessed by performing the inflation-deflation experiment $7$ times. The volunteer is placed in a supine position and instructed to relax the leg muscles. The relative Young's moduli for the $3$ different muscle groups are reconstructed using the proposed framework. The respective standard deviations of the reconstructed Young's moduli over the $7$ repetitions offer a measure for the repeatability.


\emph{Experiment 2.2: in vivo elastography during isometric knee flexion}~\\
As an application, the Young's moduli of the thigh in relaxed state are compared to those during isometric knee flexion. Knee flexion is the action of decreasing the angle between the thigh and the lower leg, essentially bending the knee. The volunteer is placed in a supine position and instructed to press their heel into the table with a constant force for the duration of the scan. Tension in the thigh builds up, but no motion results from this muscle activation (the exercise is isometric) due to the resistance from the table. The deformations in the thigh are again induced by the inflation and deflation of the pressure cuff. This experiment is conducted only once since repeated muscle activation would likely induce changes in mechanical properties, making the repeatability analysis meaningless.

\section{Results}
\subsection{In silico experiments}
\emph{Experiment 1.1: local, voxel-wise stiffness characterization}~\\
In silico Experiment 1.1 (see Fig. \ref{fig: InSilicoLoc}) is performed on a configuration that contains two distinct materials types arranged coaxially. The reconstructed spatial distribution of the Young's modulus in the noise free situation (SNR=$\infty$) is in good agreement with the ground truth elastic distribution. The small discrepancies are most likely due to partial volume effects. The addition of a small amount of noise (SNR=$100$) to the ground truth displacement fields does not visibly deteriorate the reconstructed elasticity distribution. For a larger noise level (SNR=$20$) the results are significantly worse, although the coaxial configuration can still be distinguished.

\emph{Experiment 1.2: region-wise stiffness characterization}~\\
In silico Experiment 1.2 is performed on two different elastic configurations. For the first configuration that contains two distinct materials (see Fig. \ref{fig: InSilico1}) the reconstructed Young's moduli in the noise free situation (SNR=$\infty$) are in good agreement with the ground truth values to within $0.6\%$. The addition of noise to the ground truth displacement fields does not introduce a strong bias in the solutions. The mean solution from $1000$ reconstructions with different noise realizations is within $0.5\%$, $2.2\%$ and $6.5\%$ of the ground truth relative Young's moduli for SNR=$100, 10$ and $5$ respectively. Also the standard deviations are considered reasonable for high levels of noise. 
For the second configuration that contains four distinct materials (see Fig. \ref{fig: InSilico4}) the reconstructed Young's moduli in the noise free situation (SNR=$\infty$) are in agreement with the ground truth values to within $8.6\%$. The addition of noise to the ground truth displacement fields introduces a bias in the solutions. The mean solution from $1000$ reconstructions with different noise realizations is within $8.3\%$, $10.8\%$ and $47\%$ of the ground truth relative Young's moduli for SNR=$100, 10$ and $5$ respectively. The standard deviations for this configuration are higher than for the configuration with two material types.  

\subsection{In vivo experiments}
\begin{figure*}[t!]
    \centering
    \includegraphics[width=\linewidth]{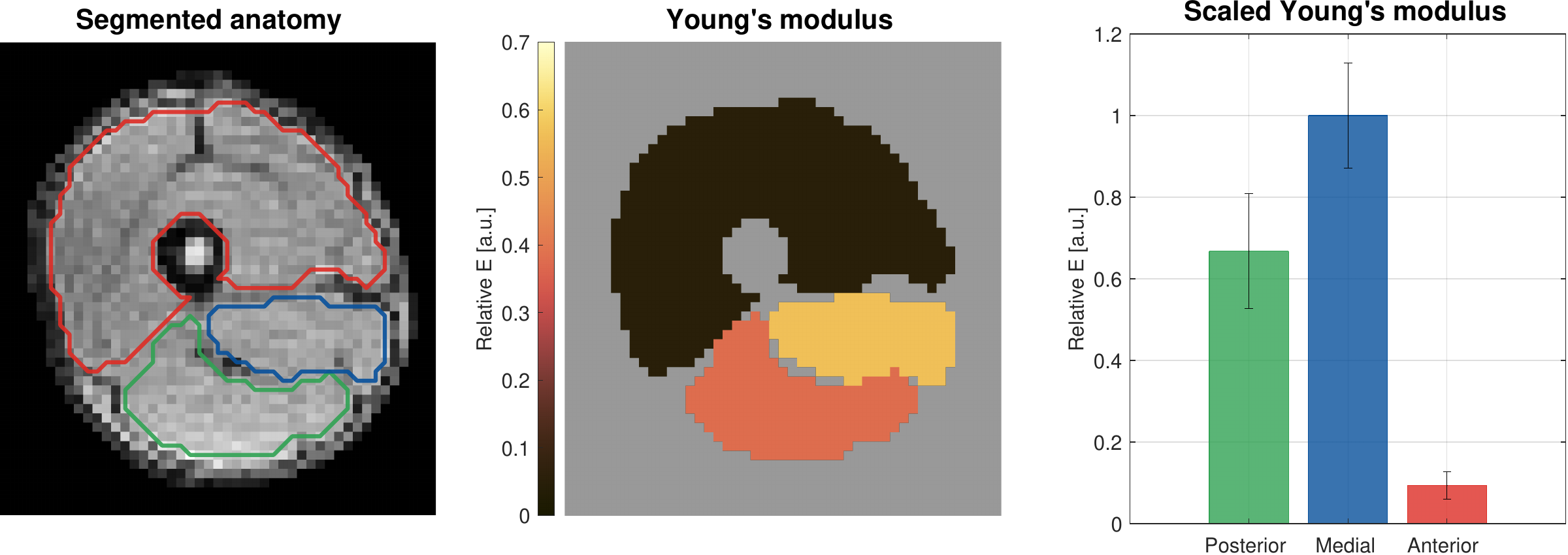}
    \caption{In vivo Experiment 2.1 investigating the in vivo repeatability on the relaxed thigh muscle. The figure on the left shows the thigh segmented into the posterior (green), medial (blue) and anterior (red) compartment. The figure in the middle shows the segmentation based spatial distribution of the relative Young's moduli reconstructed such that the sum for all materials equals $\alpha=1$. The figure on the right shows a bar chart displaying the same results, but scaled such that the relative Young's modulus of the medial is set to $1$. The error flags indicate the standard deviation from $7$ repetitions of the experiment.\\ \phantom{q}}
    \label{fig: Reproducibility}
\end{figure*}
\emph{Experiment 2.1: in vivo elastography and repeatability}~\\
The results of the in vivo repeatability study are presented in Fig. \ref{fig: Reproducibility}. These findings demonstrate that the relative Young's moduli of the posterior ($0.38$) and medial ($0.57$) compartment are significantly higher compared to the anterior compartment ($0.05$). This is in line with what we expect from physiology. In supine position, the leg is positioned such that the muscles in the back (posterior) are stretched, while the muscles in the front (anterior) remain relaxed. The passive stretching increases the muscle stiffness, making them more resistant to deformation. Reasonable standard deviations are reported, corresponding to a coefficient of variation of $21\%$, $13\%$ and $36\%$ for the posterior, medial and anterior compartments respectively, indicating good repeatability of the method in vivo.

\begin{figure}[t!]
    \centering
    \includegraphics[width=\linewidth]{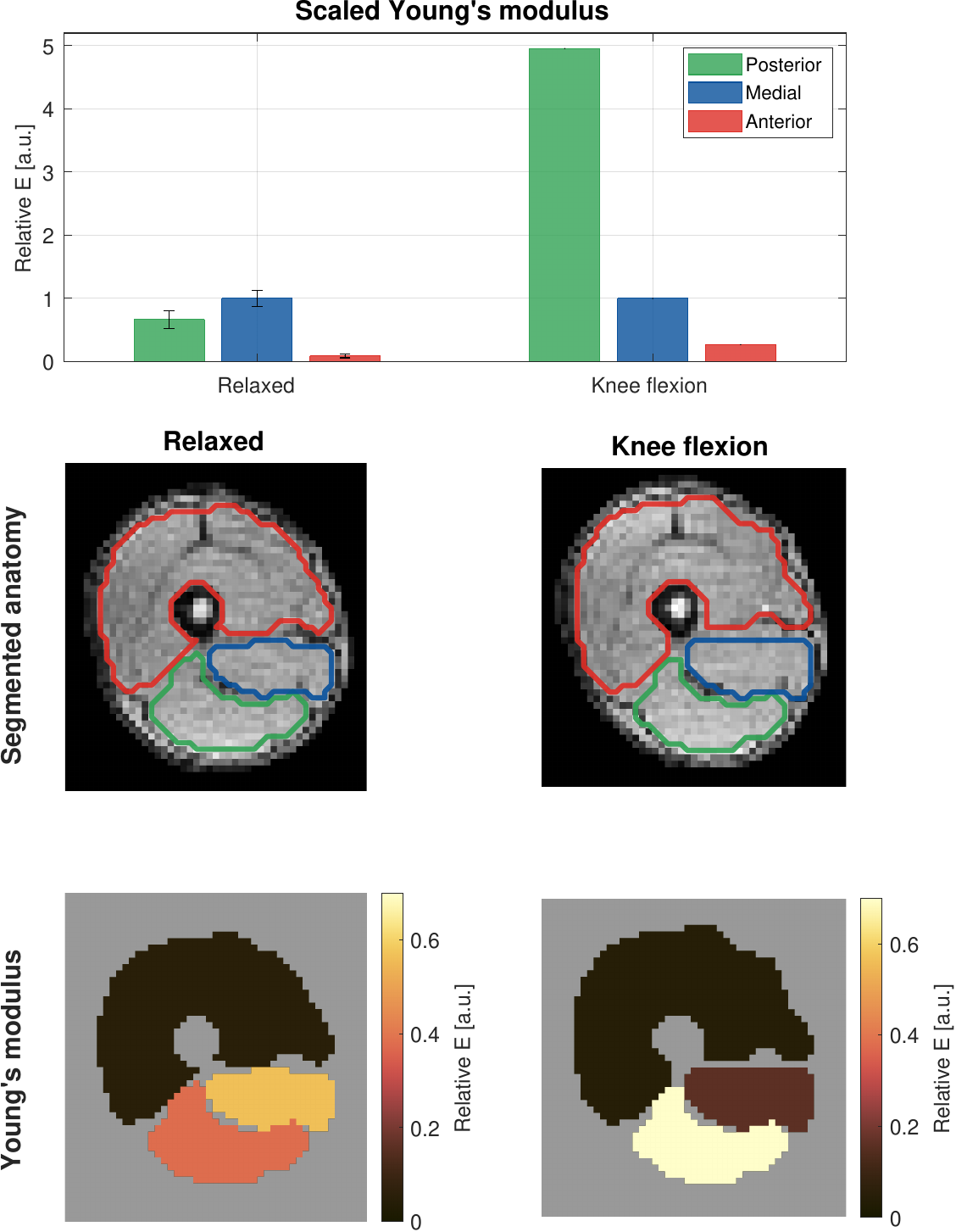}
    \caption{In vivo Experiment 2.2 comparing the relative Young's moduli of the relaxed thigh with those of the thigh under isometric knee flexion. The figure on top shows a bar chart with relative Young's moduli, scaled such that the relative Young's modulus of the medial is set to $1$. Note that the bar chart for the relaxed thigh is identical to the one in Fig. \ref{fig: Reproducibility}. The figures in the middle show the thigh in both states segmented into the posterior (green), medial (blue) and anterior (red) compartment. The figures on the bottom show the segmentation based spatial distribution of the relative Young's moduli, reconstructed such that the sum for all materials equals $\alpha=1$.}
    \label{fig: Comparison}
\end{figure}

\emph{Experiment 2.2: in vivo elastography during isometric knee flexion}~\\
The results of the application study where the relaxed thigh is compared to the thigh during isometric knee flexion (activated) are presented in Fig. \ref{fig: Comparison}. 
Since the calculated Young's moduli represent relative values, a direct comparison between the two states is not possible. However, given that the medial compartment is not involved in knee flexion, we can make an approximate comparison between the relaxed and activated state by assuming the stiffness of the medial remains the same in both conditions. In the bar chart shown in Fig. \ref{fig: Comparison}, the medial complex is normalized to $1$, facilitating direct comparison under this assumption.

\section{Discussion}\label{sec: Discussion}
In this proof-of-principle study, we introduced a noise robust reconstruction framework for in vivo quantitative assessment of stiffness properties from internal displacement fields in quasi-static conditions. Our method does not require boundary information which is generally hard to access during in vivo scanning. Furthermore, the spatial derivatives of the measured displacement fields are removed from the problem, circumventing the associated noise amplification of this operation. 
The validity of our approach was corroborated using in silico tests on a numerical phantom. Experiment 1.1, where we try to reconstruct the spatial distribution of the relative Young's modulus, indicates that for modest SNR of $20$ the results in 2D are already suboptimal. Using the region-wise homogeneity assumption based on segmentation information in Experiment 1.2 drastically improved the stability. Experiment 1.2 with two material types is most robust to noise as it has less unknowns and more displacement data available per segment compared to Experiment 1.2 with four material types. Additionally, we demonstrated the in vivo repeatability of our framework in Experiment 2.1 and, as an application, showed in Experiment 2.2 that the anterior and posterior compartment of the thigh exhibit increased stiffness during activation, as physiologically expected. During isometric knee flexion, the stiffness of the anterior and posterior compartment increases compared to the relaxed state by a factor of $2.88$ and $7.42$ respectively. The increased stiffness of the anterior compartment during muscle activation are in line with previous studies on the activation of the Vastus Lateralis where the stiffness more than doubled at $20\%$ voluntary muscle contraction\cite{bensamoun_determination_2006}. 

One limitation of this proof-of-principle work pertains to the 2D plane strain assumption. Given the complexity of the human anatomy, we can expect that the plane strain assumption might not be fulfilled exactly. 
Therefore, to account for this model imperfection and to increase the available data, an extension of the proposed reconstruction framework to 3D is a logical next step when moving to in vivo applications. Another limitation arises from the use of the isotropic elastic model, which is not optimal for describing muscles, as the presence of fibers introduces anisotropy\cite{fung_biomechanics_1993}. However, our reconstruction framework is generic in the sense that it can easily be extended to different linear elastic models. Mapping anisotropic properties using MR Elastography has already been demonstrated using fiber information from Diffusion Tensor Imaging (DTI) \cite{mcgarry_mapping_2022}. Our method possibly provides a framework for retrieving information on anisotropy without DTI, although this is speculative at this stage and should be corroborated in future research. Finally, the assumption that we are in the regime of linear elasticity should be validated in a quantitative study which is left for future work. 

From a mathematical perspective, a few remarks are warranted. First, for our inverse problem to be uniquely solvable up to a global multiplicative constant ($\boldsymbol{\mathcal{A}}$ has a one-dimensional null space), four independent displacement fields are required\cite{barbone_elastic_2004} as pointed out in section \ref{subsec: reconstruction}. Each time frame of the Spectro-dynamic displacement reconstruction is considered a distinct displacement field. However, due to the high temporal correlations, it is challenging to assess how much information is contained in the full set of $61$ displacement fields (how independent these fields are). For this reason, we would like to include displacement fields from multiple deformation strategies in the reconstruction, which is left for future work. Second, for the noise robust form of the reconstruction problem presented in \eqref{eq: inverse_problem} the solution space for the elasticity tensors $\boldsymbol{C}\in W^{1, \infty}(\Omega, \boldsymbol{T}^4_{\text{sym}})$ does not include the discontinuous ground truth that is only part of $L^{\infty}(\Omega, \boldsymbol{T}^4_{\text{sym}})$. Although this prohibits us from formulating a rigorous stability proof of the proposed method, the in silico experiments do suggest that the method is stable even for ground truth solutions outside of the solution space. For the Virtual Fields Method without boundary information, it is possible to prove $L^2$-stability of the solution in $L^{\infty}(\Omega, \boldsymbol{T}^4_{\text{sym}})$ as provided by Ammari et al. in \cite{ammari_direct_2021}. 

To improve the current framework, several extensions are proposed for future work. First, the freedom in selecting test functions $\vec{\eta}$, while adhering to the constraints outlined in section \ref{subsec: The proposed robust weak form}, can be exploited to improve the conditioning of the resulting problem matrix $\boldsymbol{\mathcal{A}}$. Second, the current knee flexion experiment uses the assumption that the stiffness of the medial compartment remains constant between the relaxed and activated states of the thigh muscle which might not be entirely accurate. To enable a more robust comparison between the two tissue states, a reference material with sufficient MRI contrast could be integrated into the setup to deform alongside the tissue. Since the mechanical parameters of this material remain unchanged between the two tissue states, it could serve as a reference for the comparison. Moreover, when the mechanical parameters of the reference material are known, the unknown global scaling of the reconstructed stiffness values can be determined. This enables us to convert the relative Young's moduli into absolute stiffness parameters. Third, when a good estimation of the elastic parameters is obtained, it can be used to regularize the Spectro-dynamic reconstruction of the displacement fields. The joint iterative reconstruction of the elastic parameters together with the time-resolved displacement fields could potentially be symbiotic; tissue elastic information can be leveraged for the estimation of the displacement fields using the conservation of momentum equation from \eqref{eq: conservation of momentum} as regularization. This should result in more accurate displacement information, which in turn benefits the reconstruction of elastic parameters. Such a joint iterative reconstruction approach has already been studied in the context of Spectro-dynamic MRI for a simple mechanical phantom \cite{heesterbeek_data-driven_2024}. 

Finally, this research can be considered as a first step in the direction of imaging elastic properties of internal organs using physiological activation. As we have access to displacement fields with a high spatiotemporal resolution and require no boundary information, our reconstruction method could allow for applications such as the assessment of cardiac stiffness using the internal actuation from the heart contractions or stiffness analysis of the liver using the motion induced by cardiac and respiratory activity. 

\section{Conclusion}
A noise robust reconstruction framework for tissue elastic parameter estimation from quasi-static displacement information is proposed and successfully validated. The reconstruction framework does not require boundary information and avoids spatial derivatives that amplify noise. In silico evaluations using a numerical phantom demonstrate the noise robustness of our framework, while in vivo experiments on the thigh muscle indicate good repeatability and yield physiologically meaningful results.


\bibliographystyle{unsrt}
\bibliography{references.bib}

\end{document}